%% file: main.tex
\pdfoutput=1

\documentclass{article}

\usepackage[preprint]{neurips_2024}

\usepackage[utf8]{inputenc}
\usepackage[T1]{fontenc}
\usepackage{lmodern}        
\usepackage{hyperref}
\usepackage{url}
\usepackage{booktabs}
\usepackage{amsmath, amssymb}
\usepackage{graphicx}
\usepackage{xcolor}
\usepackage{multirow}
\usepackage{array}
\usepackage{caption}
\usepackage{subcaption}
\usepackage{microtype}
\usepackage{longtable}      
\providecommand{\tightlist}{%
  \setlength{\itemsep}{0pt}\setlength{\parskip}{0pt}}  
\usepackage{fancyvrb}       
\usepackage{fvextra}        
\usepackage{tikz}           
\usetikzlibrary{positioning,arrows.meta,fit,backgrounds,calc}

\hypersetup{
  colorlinks=true,
  linkcolor=blue!50!black,
  citecolor=blue!50!black,
  urlcolor=blue!50!black
}

\title{Industry Classification of GitHub Repositories\\
Using the North American Industry Classification System (NAICS)}

\author{%
  Kevin Xu \\
  GitHub \\
  \texttt{khxu@github.com}
  \And
  Alexander Quispe\thanks{Corresponding author.} \\
  GitHub \\
  \texttt{alexanderquispe@github.com}
}

\date{}

\begin{document}

\maketitle

\begin{abstract}
GitHub hosts hundreds of millions of public repositories, but the
platform exposes no native mapping from repositories to standardized
industry sectors. This gap limits empirical work on the geography of
innovation, the industrial composition of open-source production, and
the diffusion of new technologies across economic sectors. We present
\textbf{NAICS-GH}, a publicly released corpus of \textbf{6{,}588 GitHub
repositories} drawn from source pools covering the United States, the
European Union, and Australia, each labeled with a 2-digit sector from
the North American Industry Classification System (NAICS 2022). Labels
are produced by a retrieve-and-verify pipeline that combines
BAAI/bge-large-en embeddings, FAISS retrieval, and GPT-4.1 rubric
scoring. The pipeline narrows about 1.37 million source repositories to
\textbf{31{,}178 candidate repository-sector pairs} and retains
\textbf{6{,}588 high-confidence labels} with score at least 8.
Re-running the retrieval pipeline end to end reproduces the candidate
set to within 0.03 percent. On a 2{,}421-repository human-validated
random sample, the released labels attain \textbf{96.98 percent
precision}, with Wilson 95 percent confidence interval [96.23, 97.59].
We benchmark six pretrained encoders on the released corpus;
RoBERTa-large reaches \textbf{86.45 percent F1} and \textbf{86.35
percent accuracy} on a held-out 20 percent test set. The dataset,
Croissant metadata, pipeline code, prompts, and fine-tuned checkpoint
are released under CC-BY-4.0 and MIT licenses.
\end{abstract}

\section{Introduction}
\label{sec:intro}

GitHub hosts tens of millions of public repositories, but the platform
provides no native indication of which \emph{industry} a repository
serves. Knowing whether a project is fintech, agritech, healthcare
software, or educational tooling matters to policy makers tracking the
geography of innovation, to companies measuring open-source adoption, and
to economists studying the labor and capital allocation of the software
sector. We address this gap by releasing the first multi-region,
publicly available corpus of GitHub repositories labeled with NAICS---the
industry-classification standard used by the United States, Canadian, and
Mexican statistical agencies.

\paragraph{Contributions.}
\begin{itemize}
  \item \textbf{Dataset.} We release NAICS-GH, \textbf{6{,}588} GitHub
    repositories from the USA, EU, and Australia labeled with 2-digit
    NAICS codes, alongside the full pipeline outputs (sector, score,
    rationale, repository URL).
  \item \textbf{Pipeline.} A reproducible two-stage retrieve-and-verify
    labeling pipeline (BAAI/bge-large-en embeddings + FAISS
    retrieval, followed by GPT-4.1 rubric scoring) suitable for
    adapting to other industry taxonomies. The full pipeline narrows
    $\sim$1.37M source repositories to \textbf{31{,}178 candidate
    pairs} via retrieval, then to \textbf{6{,}588 high-confidence
    labels} via LLM verification at score $\geq 8$. An end-to-end
    re-run reproduces the candidate set to within $\pm 0.03\%$.
  \item \textbf{Validation.} A 2{,}421-repository manually re-checked
    gold subset, confirming 96.98\% label precision overall and
    monotonically increasing precision as the GPT score rises from 8 to
    10.
  \item \textbf{Benchmark.} A head-to-head comparison of six
    pretrained encoders (RoBERTa, ModernBERT, DeBERTa-v3 in base and
    large sizes); RoBERTa-large is strongest at 86.45\% test F1, and
    the fine-tuned checkpoint is available on the Hugging Face Hub.
\end{itemize}

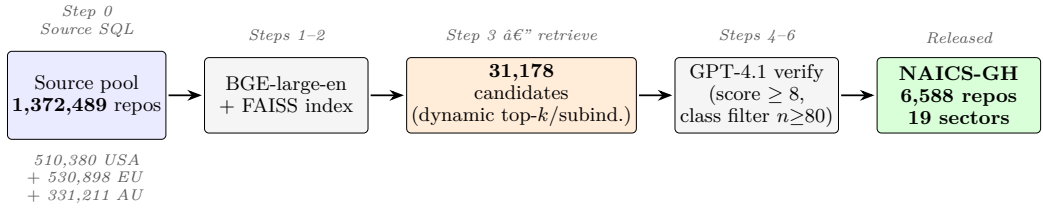
\begin{figure}[ht]
\centering
\resizebox{\linewidth}{!}{%
\begin{tikzpicture}[
    >=Stealth,
    font=\footnotesize,
    node distance=2.5mm and 6mm,
    box/.style={draw, rounded corners=2pt, align=center,
                minimum width=22mm, minimum height=10mm,
                inner sep=2pt},
    src/.style={box, fill=blue!8, minimum width=18mm, minimum height=8mm},
    proc/.style={box, fill=gray!8, minimum width=26mm, minimum height=12mm},
    cand/.style={box, fill=orange!15, minimum width=26mm, minimum height=12mm},
    result/.style={box, fill=green!15, font=\footnotesize\bfseries,
                   minimum width=26mm, minimum height=12mm},
    note/.style={font=\scriptsize\itshape, gray!70!black, align=center},
    arr/.style={->, thick, shorten <=1pt, shorten >=1pt}
]

\node[src, minimum width=24mm, minimum height=14mm]
                                    (pool) {Source pool\\\textbf{1{,}372{,}489} repos};
\node[note, above=1mm of pool]      {Step 0\\Source SQL};
\node[note, below=1mm of pool, text width=30mm]
    {510{,}380 USA\\$+$ 530{,}898 EU $+$ 331{,}211 AU};

\node[proc, right=of pool]          (bge) {BGE-large-en\\$+$ FAISS index};
\node[note, above=1mm of bge]       {Steps 1--2};

\node[cand, right=of bge]           (retr)
    {\textbf{31{,}178}\\candidates\\(dynamic top-$k$/subind.)};
\node[note, above=1mm of retr]      {Step 3 â€” retrieve};

\node[proc, right=of retr]          (gpt)
    {GPT-4.1 verify\\(score $\geq 8$,\\class filter $n{\geq}80$)};
\node[note, above=1mm of gpt]       {Steps 4--6};

\node[result, right=of gpt]         (rel)
    {NAICS-GH\\6{,}588 repos\\19 sectors};
\node[note, above=1mm of rel]       {Released};

\draw[arr] (pool.east) -- (bge.west);
\draw[arr] (bge.east)  -- (retr.west);
\draw[arr] (retr.east) -- (gpt.west);
\draw[arr] (gpt.east)  -- (rel.west);

\end{tikzpicture}%
}
\caption{NAICS-GH end-to-end pipeline. A Presto/Trino SQL extraction
from GitHub's data warehouse yields a source pool of
1{,}372{,}489 public repositories across three jurisdictional
extractions (510{,}380 USA + 530{,}898 EU + 331{,}211 AU), each
processed by an identical pipeline configuration; per-row country
attributes are not retained in any released artifact.
BAAI/bge-large-en embeddings indexed in FAISS retrieve the top-$k$
most semantically similar repositories per NAICS subindustry query
($k = \max(20, \lceil 400/n \rceil)$ for a sector with $n$
subindustries), producing \textbf{31{,}178 unique candidate
(repository, sector) pairs} after per-sector deduplication
(10{,}983 USA + 10{,}531 EU + 9{,}664 AU). GPT-4.1
(snapshot \texttt{gpt-4.1-2025-04-14}) scores each candidate against a
structured rubric; we retain repositories with score $\geq 8$ and
apply a minimum-class-size filter ($n{\geq}80$ per sector) to drop
sector 55, yielding the released \textbf{NAICS-GH} corpus of
\textbf{6{,}588 repositories} across 19 of the 20 NAICS 2-digit
sectors.}
\label{fig:pipeline}
\end{figure}

\paragraph{Pipeline at a glance.} Figure~\ref{fig:pipeline} summarizes
the end-to-end retrieve-and-verify pipeline used to construct
NAICS-GH from raw public-repository data.

\section{Related Work}
\label{sec:related}

\paragraph{GitHub repository classification.}
GitHub repositories have long lacked standardized industry labels, and
existing work has approached this gap from three angles. The first
relies on user-declared topic tags as the label space:
\citet{zhang2019higitclass} use a keyword-driven hierarchical model,
\citet{izadi2020topic} cast it as a multi-label problem, and
\citet{sas2022gitranking} rank GitHub topics by relevance with active
sampling. The second labels repositories by software application
domain: \citet{zanartu2022categorizing} build a 5{,}000-repository,
five-domain classifier achieving $\sim 70\%$ precision, and
\citet{balla2026dragon} scale to \textbf{825{,}000 repositories} with
ground-truth topics from the Software Heritage archive, reporting
F1$@5 = 60.8\%$ with file-tree and README features. None of these
works adopt a standardized economic taxonomy or use a modern LLM
verifier. NAICS-GH differs from each on all three axes:
(i)~we adopt the NAICS 2-digit hierarchy, (ii)~we use GPT-4.1
verification on top of dense-retrieval candidates, and (iii)~we
cover three jurisdictions (USA, EU, AU) rather than a single one.

\paragraph{NAICS classification of text.}
The United States Census Bureau uses NAICS to label every business
establishment, but published machine-learning tools for the
taxonomy are scarce. The closest prior work is BEACON
\citep{dumbacher2025beacon}, a text-classification system deployed in
the 2022 U.S. Economic Census that helps respondents self-classify
their business activity via NLP, machine learning, and information
retrieval; BEACON was used over half a million times during the 2022
Census. Two structural differences distinguish it from our work:
BEACON's unit of analysis is a business establishment described by
its owner (we classify GitHub repositories described by community
metadata), and BEACON's algorithm and training data are not publicly
released (we release both the labeled corpus and the pipeline code).
To our knowledge, NAICS-GH is the first publicly available
NAICS-labeled corpus of software repositories.

\paragraph{LLM-as-labeler and weak supervision.}
Our retrieve-and-verify pipeline sits in the weak-supervision
tradition where heuristic or programmatic functions assign noisy
labels that are then aggregated and refined
\citep{ratner2016dataprogramming,ratner2017snorkel}. LLMs have
recently emerged as a powerful source of such labels:
\citet{gilardi2023chatgpt} report that ChatGPT exceeds crowd-workers
in accuracy and agreement on text-annotation tasks at a fraction of
the cost, and \citet{elumar2025cost} develop cost-aware majority
voting across multiple LLMs to mitigate individual-model bias.
NAICS-GH instantiates this tradition with a specific domain
(software repositories), a specific taxonomy (NAICS), and a
verification step that uses one high-capability model (GPT-4.1) with
a structured rubric and human re-validation on a stratified
subsample.

\paragraph{LLMs applied to GitHub content.}
A growing body of work uses LLMs to classify or extract structured
data from the GitHub platform, but with different units of analysis.
\citet{mehmood2025llm} fine-tune BERT, RoBERTa, and DistilBERT (with
LoRA) to classify the \emph{sections} of a README into eight
structural roles (What/Why/How/$\ldots$), reaching F1 = 0.98 on the
4{,}226-section \citet{zanartu2022categorizing}-derived benchmark.
\citet{chen2025forge} use an LLM-driven framework to construct a
smart-contract vulnerability dataset at scale, and
\citet{joynab2026threads} propose a multi-LLM pipeline that extracts
community knowledge from GitHub issue discussions. Our work is
complementary to these: we use the README alongside the description
and topics to classify the \emph{whole repository} into a 19-class
industry taxonomy, an output space that is broader and more
semantically overlapping than the section- or vulnerability-typed
analogs above.

\paragraph{Datasheets.} We provide a Datasheet for
Datasets~\citep{gebru2021datasheets} in Appendix~\ref{app:datasheet}.

\section{Dataset Construction}
\label{sec:construction}

\subsection{Source repositories}
\label{sec:source}

We extract source repositories from GitHub's internal Trino-on-Hive data
warehouse via a Presto/Trino SQL query against the
\texttt{hive.canonical.repositories\_current},
\texttt{hive.canonical.accounts\_current},
\texttt{hive.suez.readme\_current}, and
\texttt{delta.snapshots.github\_collab\_commit\_contributions} tables.
The query was executed against data current through \textbf{April 15,
2025} (USA presample; EU readme extraction August 15, 2025). The full
SQL is in Appendix~\ref{app:sql}.

\paragraph{Inclusion criteria.} A repository is included in the source
pool if it satisfies all of the following:
\begin{itemize}
  \item Public, non-fork, non-spammy-owner GitHub repository
    (\texttt{is\_public = TRUE}, \texttt{is\_fork = FALSE},
    \texttt{NOT is\_spammy\_owner}).
  \item At least one star (\texttt{num\_stars $\geq 1$}).
  \item Declares a top-level \texttt{README.md} (case-insensitive)
    of at least 750 bytes.
  \item Disk usage greater than zero, and non-empty repository
    description.
  \item At least 6 total commits and at least 2 distinct commit
    contributors
    (\texttt{HAVING SUM(commit\_count) > 5} and
    \texttt{COUNT(DISTINCT user\_id) $\geq 2$}).
  \item Owner-account country code in the per-region list
    (\S\ref{sec:source:regions}).
\end{itemize}

\paragraph{Jurisdiction split.}
\label{sec:source:regions}
Repositories are assigned to a region by the
\texttt{country\_account} field of the \emph{owner account} in
\texttt{accounts\_current}, joined via
\texttt{owner\_dotcom\_id} (not by top-contributor location or geo-IP):
\begin{itemize}
  \item \textbf{USA}: \texttt{'US'} (510{,}380 repositories).
  \item \textbf{EU} (27 codes, EU-27 excluding the UK):
    \texttt{AT, BE, BG, CY, CZ, DE, DK, EE, ES, FI, FR, GR, HR, HU,
    IE, IT, LT, LU, LV, MT, NL, PL, PT, RO, SE, SI, SK}
    (530{,}898 repositories).
  \item \textbf{AU}: \texttt{'AU'} (331{,}211 repositories).
\end{itemize}

\paragraph{Total source pool.} 1{,}372{,}489 repositories across the
three jurisdictions. Each repository carries its owner-and-name
identifier (\texttt{nwo}), description, topic tags, README content,
SPDX license identifier, and activity metadata (stars, commits,
contributors, issue counts, last-commit timestamp).

\paragraph{Per-region processing, country-blind release.}
\label{sec:source:merge}
The source SQL emits three jurisdictional extractions, and each
subsequent generation stage (embedding, retrieval, LLM verification,
score filtering) runs \emph{per region} with an identical
configuration; the three score-filtered outputs are then concatenated
into the released corpus (\S\ref{sec:verification}). No released
artifact retains a per-row \texttt{country\_code} column or any other
per-row jurisdictional attribute: rows are shuffled at assembly so
regional provenance is not recoverable from row position, and
jurisdiction appears in this paper only as aggregate statistics. To
enable external reproduction without access to jurisdiction-tagged
data, we additionally publish a jurisdiction-blind variant of the
pipeline that operates on the concatenation of the three extractions
as a single corpus (\S\ref{sec:release}).

\subsection{NAICS sector taxonomy}
\label{sec:naics-taxonomy}

We use a pre-built JSON taxonomy file
(\texttt{naics\_titles\_by\_group\_6digit\_clean.json}) with 20 entries,
one per 2-digit NAICS sector code. Each entry is a single string of
subindustry titles separated by semicolons, sourced from the 6-digit
NAICS hierarchy and grouped by their parent 2-digit code. We consume
the file as-is; no 6-to-2-digit collapsing is performed by our
pipeline.

Across the 20 sectors there are \textbf{1{,}029 distinct subindustry
phrases} in total. The count per sector varies widely: Manufacturing
(31--33) has 346 subindustries, the most by far; Utilities (22),
Accommodation/Food (72), and Educational Services (61) have the
fewest (14, 15, and 17 respectively). The full taxonomy is in
Appendix~\ref{app:naics}.

The taxonomy contributes to the pipeline in two distinct ways:
\begin{itemize}
  \item \emph{Semantic retrieval (\S\ref{sec:retrieval})} uses the
    \textbf{individual subindustry phrases} as queries, one query per
    phrase.
  \item \emph{LLM verification (\S\ref{sec:verification})} uses the
    \textbf{full concatenated string} for each sector as the rubric
    anchor presented to GPT-4.1 alongside the candidate README.
\end{itemize}

\subsection{Semantic retrieval}
\label{sec:retrieval}

\paragraph{Embedding the source corpus.}
For each region, we encode the first 1{,}000 characters of each
repository's raw README using
\textbf{BAAI/bge-large-en}~\citep{xiao2023cpack}, with the asymmetric
BGE prefixes: \texttt{"Represent this document for retrieval: "} for
documents and \texttt{"Represent this query for retrieval: "} for
queries. (Repository description and topic tags enter the pipeline at
the verification stage, \S\ref{sec:verification}, where the LLM
receives the full concatenated text.) Embeddings are 1024-dimensional
and L2-normalized; inference uses half-precision (FP16) on GPU with a
batch size of 1024. We build an exact inner-product FAISS index
(\texttt{IndexFlatIP})~\citep{johnson2021faiss}, which under
normalization computes cosine similarity.

\paragraph{Per-subindustry query loop.}
Retrieval is driven by the NAICS taxonomy loaded in
\S\ref{sec:naics-taxonomy}. For each of the 20 sectors we split the
sector's semicolon-separated string into its constituent subindustry
phrases and issue \emph{one FAISS query per phrase} with the template
\texttt{"Repositories about \{subindustry\}"}. The number of nearest
neighbors retrieved per query, \texttt{effective\_k}, is set
adaptively:

\begin{equation*}
  \mathtt{effective\_k} \;=\;
  \begin{cases}
    20 & \text{if } n > 20 \\
    \max\!\left(20,\;\lceil 400 / n \rceil\right) & \text{if } n \leq 20
  \end{cases}
\end{equation*}

where $n$ is the number of subindustry phrases in the current sector.
For the 17 sectors with $n > 20$ this gives a uniform $k = 20$; for
the three sectors with narrow taxonomies the formula boosts $k$ so
that each sector retrieves on the order of 400 candidates regardless
of $n$: \texttt{effective\_k} $= 29$ for Utilities ($n=14$), $27$ for
Accommodation/Food ($n=15$), and $24$ for Educational Services
($n=17$).

\paragraph{Tagging and deduplication.}
Each returned row is annotated with the sector code, the specific
subindustry phrase that produced the match, and the BGE cosine
similarity. The loop across all 1{,}029 subindustry queries retrieves
exactly 20{,}879 rows per region ($\sum_s n_s \times k_s$, overlap
allowed); deduplicating on \texttt{(repository, sector)} within each
region yields \textbf{31{,}178 unique candidate (repository, sector)
pairs} (10{,}983 USA + 10{,}531 EU + 9{,}664 AU).

\paragraph{Why this design.}
The retrieval is intentionally over-inclusive --- the BGE cosine
similarities of the returned candidates concentrate in a narrow
$[0.78, 0.89]$ band, so the score alone cannot distinguish true
matches from near-matches. The expensive verification step
(\S\ref{sec:verification}) handles that discrimination. Querying per
subindustry rather than per sector lets us cast a wider net (e.g.,
Agriculture is queried 64 times rather than once) and exposes
downstream code to the specific subindustry that surfaced each
candidate, which is useful for error analysis.

\subsection{LLM verification}
\label{sec:verification}

We score each candidate pair using \textbf{GPT-4.1} (model snapshot
\texttt{gpt-4.1-2025-04-14}) via the GitHub Copilot LLM-lab endpoint
(\texttt{api-model-lab.githubcopilot.com/chat/completions}) with a
structured prompt (Appendix~\ref{app:prompt}).

\paragraph{What the model sees.}
For each candidate \texttt{(repo, sector)} pair, the API call sends two
messages:
\begin{itemize}
  \item A \emph{system message} identifying the model as ``a domain
    expert in economic classification systems with a focus on NAICS
    industry \texttt{\{code\}}.''
  \item A \emph{user message} containing the structured prompt. The
    prompt includes:
    \begin{itemize}
      \item The repository's \textbf{combined text} --- not the bare
        README, but the concatenation of the repository's description,
        topic tags, and cleaned README, joined with explicit field
        labels:
        \begin{center}
        \small
        \texttt{description: \{desc\}, topics: \{topics\}, readme: \{readme\}}
        \end{center}
        This composite string is what is inserted inside the prompt's
        \texttt{<readme>\ldots</readme>} tags. The repository name is
        not included.
      \item Preprocessing: Markdown code blocks, inline code, images,
        links, HTML tags, and excess whitespace are stripped, and the
        result is truncated to \textbf{3{,}000 whitespace-separated
        tokens} (i.e., a word-count proxy, not a BPE token count).
      \item The candidate NAICS sector code and the full concatenated
        subindustry definition from the taxonomy
        (\S\ref{sec:naics-taxonomy}).
      \item A four-criterion rubric (industry-specific software,
        sector-relevant functionality, industry-domain applications,
        sector-specific data/research) and a 1--10 scoring guide.
      \item Instructions to reply as a JSON object nested under the
        outer key \texttt{"NAICS \{code\}"} with string-valued
        \texttt{match} (\texttt{"Yes"} or \texttt{"No"}),
        string-valued \texttt{score} (\texttt{"1"}--\texttt{"10"}),
        and free-form \texttt{rationale}. The downstream pipeline
        coerces \texttt{match} to a boolean and \texttt{score} to an
        integer.
    \end{itemize}
\end{itemize}

\paragraph{API parameters.}
We call GPT-4.1 with \texttt{temperature = 0} for near-deterministic
outputs, and set \texttt{max\_tokens} dynamically as
$\min(3500,\;128000 - \mathrm{input\_tokens})$ based on the remaining
context budget (input tokens are counted with
\texttt{tiktoken.encoding\_for\_model("gpt-4.1")}, falling back to
\texttt{cl100k\_base}). On non-200 API responses we retry up to five
times with linear backoff (\texttt{sleep = 5} $\times$ attempt
seconds). Replies are parsed by extracting the JSON object via regular
expression and calling \texttt{json.loads}; a small fallback handles
the case when the model wraps its JSON in Markdown fences.

\paragraph{Filtering.}
We retain repositories whose returned \texttt{score} is at least 8.

\subsection{Filtering and final assembly}
We concatenate the three regional score-$\geq 8$ outputs and apply a
minimum-samples filter of 80 repositories per NAICS sector,
eliminating Sector~55 (``Management of Companies and Enterprises''),
which had only 13 repositories. After deduplication and cleanup we
obtain the released NAICS-GH training corpus of
\textbf{6{,}588 repositories} spanning 19 of the 20 NAICS sectors.

\paragraph{Pipeline funnel.}
Table~\ref{tab:funnel} summarizes the end-to-end retention from raw
source to released training corpus.

\begin{table}[ht]
\centering
\caption{End-to-end retention from the GitHub warehouse extraction to
the released NAICS-GH training corpus, by jurisdiction.}
\label{tab:funnel}
\small
\begin{tabular}{lrrrr}
\toprule
Stage & USA & EU & AU & Total \\
\midrule
Source pool (\S\ref{sec:source}) & 510{,}380 & 530{,}898 & 331{,}211 & 1{,}372{,}489 \\
Step 3: raw retrieval ($\sum_s n_s k_s$, overlap allowed) & 20{,}879 & 20{,}879 & 20{,}879 & 62{,}637 \\
Step 3: unique candidates (\S\ref{sec:retrieval}) & 10{,}983 & 10{,}531 & 9{,}664 & 31{,}178 \\
Step 4: GPT-4.1 score $\geq$ 8 & 2{,}529 & 2{,}039 & 2{,}021 & 6{,}589 \\
Step 5--6: concat $+$ class filter & --- & --- & --- & \textbf{6{,}588} \\
\bottomrule
\end{tabular}
\end{table}

Overall, only \textbf{0.48\%} of the source pool ends up labeled in
the released corpus, and \textbf{21.1\%} of FAISS-retrieved unique
candidates survive GPT-4.1 verification at the score-$\geq 8$
threshold.

\section{Dataset Description}
\label{sec:description}

The released NAICS-GH corpus is a single
\textbf{parquet file with 6{,}588 rows and 6 columns}
(\texttt{train\_data\_gpt\_ab8\_score\_with\_code.parquet}, 12 MB on
disk). Every cell is non-null. The dataset is published as a single
training corpus; downstream code (\S\ref{sec:benchmark}) applies a
stratified $70 / 10 / 20$ split with \texttt{seed=42} to produce
train, validation, and test sets of 4{,}611, 659, and 1{,}318 rows
respectively.

\subsection{Schema}

\begin{table}[ht]
\centering
\caption{Released schema of NAICS-GH
(\texttt{train\_data\_gpt\_ab8\_score\_with\_code.parquet}).}
\label{tab:schema}
\small
\begin{tabular}{lll}
\toprule
Column & Type & Description \\
\midrule
\texttt{name\_repo} & string & Repository short name (no owner prefix). \\
\texttt{description} & string & Repository description from GitHub. \\
\texttt{topics} & string & Semicolon-joined topic tags; empty if none. \\
\texttt{readme\_content} & string & Cleaned README text. \\
\texttt{label} & int64 & Integer class encoding $0\ldots 18$. \\
\texttt{code} & string & 2-digit NAICS sector code (string form). \\
\bottomrule
\end{tabular}
\end{table}

The columns \texttt{nwo}, \texttt{match}, \texttt{score},
\texttt{rationale}, and \texttt{repo\_url} present in upstream
intermediate files are intentionally dropped from the release so
that the corpus is model-ready (no LLM-provenance, no PII via
repo-owner URLs). The mapping between \texttt{label} and \texttt{code}
is monotonic in NAICS code order:
$0 \!\to\! 11,\, 1 \!\to\! 21,\, 2 \!\to\! 22,\, 3 \!\to\! 23,\, 4 \!\to\! 31\text{-}33,\, \ldots,\, 18 \!\to\! 92$.

\subsection{Size and sector coverage}
\label{sec:desc:size}

The released corpus contains \textbf{6{,}588 repositories} spanning
\textbf{19 of the 20 NAICS sectors}. Sector 55, ``Management of
Companies and Enterprises,'' contained only 13 candidates that passed
GPT-4.1 verification at score $\geq 8$ --- below our minimum
class-size threshold of 80 --- and is therefore absent. All 19
labels in $\{0,\ldots,18\}$ are represented in the file.

\paragraph{Sector counts and imbalance.}
Per-sector counts are given in Table~\ref{tab:sector-counts}. The
distribution is moderately imbalanced: the smallest retained sector
is 23 (Construction) with \textbf{82} repositories and the largest is
44--45 (Retail Trade) with \textbf{641}, an imbalance ratio of
$\approx 7.82\times$. The mean per-sector count is 346.7 with
standard deviation 150.8; the median is 372. Twelve sectors fall
within $\pm 25\%$ of the mean, and the lower tail (sectors 21
Mining, 23 Construction, 42 Wholesale, 56 Admin/Waste) drives most
of the imbalance.

\begin{table}[ht]
\centering
\caption{Per-sector counts in the released NAICS-GH training corpus
($n=6{,}588$) drawn from
\texttt{train\_data\_gpt\_ab8\_score\_with\_code.parquet}. Sector 55
(``Management of Companies and Enterprises'') is not present because
only 13 candidates passed GPT-4.1 verification, below the 80-sample
minimum-class-size threshold.}
\label{tab:sector-counts}
\small
\input{validation/output/sector_counts_table.tex}
\end{table}


\subsection{Text characteristics}
\label{sec:desc:text}

Tables~\ref{tab:text-stats-chars} and \ref{tab:text-stats-words} give
character-length and whitespace-token-length statistics, respectively.
The README field dominates the input by a wide margin: its median
length is 1{,}912 characters (275 words) and its 99th percentile is
roughly 3{,}500 words --- comparable to the 3{,}000-word truncation
limit used by the LLM verification step (\S\ref{sec:verification}).
Only \textbf{$\sim$1.5\% of rows} (100 of 6{,}588) have a combined
text length that would have been truncated.

\begin{table}[ht]
\centering
\caption{Character-length statistics of the four text fields in the
released corpus.}
\label{tab:text-stats-chars}
\small
\begin{tabular}{lrrrr}
\toprule
Field & min & median & mean & max \\
\midrule
\texttt{name\_repo} & 2 & 14 & 16 & 85 \\
\texttt{description} & 2 & 65 & 88 & 1{,}626 \\
\texttt{topics} & 0 & 0 & 21 & 384 \\
\texttt{readme\_content} & 3 & 1{,}912 & 3{,}387 & 212{,}122 \\
\bottomrule
\end{tabular}
\end{table}

\begin{table}[ht]
\centering
\caption{Whitespace-tokenized word-count statistics for
\texttt{description} and \texttt{readme\_content}.}
\label{tab:text-stats-words}
\small
\begin{tabular}{lrrrrrr}
\toprule
Field & min & p25 & median & p75 & p99 & max \\
\midrule
\texttt{description} & 1 & 6 & 10 & 16 & 52 & 237 \\
\texttt{readme\_content} & 1 & 153 & 275 & 513 & 3{,}557 & 38{,}576 \\
\bottomrule
\end{tabular}
\end{table}

\paragraph{README size distribution.}
The distribution of README character lengths in the released corpus
is roughly log-normal:

\begin{center}
\small
\begin{tabular}{lrlr}
\toprule
README size (chars) & Rows & README size (chars) & Rows \\
\midrule
$<$ 500 & 340 & 2{,}000--5{,}000 & 2{,}121 \\
500--1{,}000 & 1{,}135 & 5{,}000--10{,}000 & 704 \\
1{,}000--2{,}000 & 1{,}964 & $>$ 10{,}000 & 324 \\
\bottomrule
\end{tabular}
\end{center}

About 76\% of READMEs are in the 500--5{,}000-character range; only
$\approx 5\%$ exceed 10{,}000 characters. The lower bound (READMEs
$<$ 500 characters, $\approx 5\%$ of the file) reflects the source
SQL's $\geq 750$-byte threshold combined with downstream Markdown
cleanup (code blocks, HTML, URLs stripped) which removes some bytes
before the field is stored.

\paragraph{Topics field.}
The \texttt{topics} field is empty in \textbf{4{,}533 of 6{,}588 rows
($\approx 68.8\%$)}. Among the 2{,}055 rows that have any tags
declared, the median is 5 tags per row (mean 6.0, p75 = 7, p99 =
20, max = 20). The field is stored as a semicolon-joined string
(e.g., \texttt{"oscibio;\,lifewatch;\,biologging;\,r-package"}).
Downstream code should treat empty strings as ``no tags declared,''
not as missing data, and should not assume the field is consistently
populated.

\subsection{Known duplicates}
\label{sec:desc:dup}

Because the canonical \texttt{(owner/repo)} identifier (\texttt{nwo})
is dropped from the release, some short-name collisions appear in the
file:

\begin{itemize}
  \item \textbf{6{,}141 unique \texttt{name\_repo} values across
    6{,}588 rows} --- i.e., \textbf{447 rows share their
    \texttt{name\_repo} with at least one other row}.
  \item Of those, \textbf{113 \texttt{(name\_repo, code)} pairs are
    duplicated}: two or more distinct repositories sharing the same
    short name independently passed verification into the same NAICS
    sector.
\end{itemize}

These are not accidental duplicates of the same repository --- they
are distinct repositories with the same short name (e.g., two
different organizations both maintaining a repository called
\texttt{api}, \texttt{docs}, or \texttt{awesome}). Users joining
external metadata by \texttt{name\_repo} alone should expect
ambiguity; the upstream intermediate files produced in
\S\ref{sec:construction} retain \texttt{nwo} as a globally unique
identifier for users who need an unambiguous join.

\subsection{Score distribution}
\label{sec:desc:score}


\section{Validation}
\label{sec:validation}

\subsection{Gold-set construction}
Research assistants reviewed a random sample of \textbf{2{,}421
repositories} drawn from the GPT-4.1-labeled output. For each
repository they inspected the GitHub page and judged whether the
GPT-assigned NAICS sector was \emph{Correct} or \emph{Incorrect}
given the criteria of Section~\ref{sec:verification}. Each row was
reviewed by a single annotator; no inter-annotator agreement
coefficient was computed (this is acknowledged as a limitation in
\S\ref{sec:limitations}).

\subsection{Headline result}
\textbf{GPT-4.1 label precision: 96.98\%} ($2{,}348$ of $2{,}421$
gold rows judged correct; 95\% Wilson confidence interval
$[96.23\%, 97.59\%]$).

\subsection{Precision by sector}
Table~\ref{tab:per-sector-precision} reports per-sector precision
in the gold sample with 95\% Wilson intervals. Six sectors
(21 Mining, 48--49 Transportation, 51 Information, 53 Real Estate,
71 Arts \& Entertainment, 92 Public Administration) reach
\textbf{100\% precision}. Two sectors fall noticeably below the
overall precision: \textbf{31--33 Manufacturing (73.0\%)} and
\textbf{42 Wholesale Trade (73.0\%)}, indicating that the LLM rubric
struggles to separate truly sector-specific manufacturing or
wholesale software from generic operations and supply-chain tooling.

\begin{table}[ht]
\centering
\caption{Per-sector precision in the gold sample
($n=2{,}421$), sorted descending. Wilson 95\% intervals.}
\label{tab:per-sector-precision}
\small
\input{validation/output/per_sector_precision_table.tex}
\end{table}

\subsection{Precision rises monotonically with GPT score}
The score returned by GPT-4.1 correlates with human-judged
correctness in the expected direction: precision rises from
\textbf{90.76\%} at score $8$ to \textbf{97.81\%} at score $9$ and
\textbf{99.30\%} at score $10$ (Table~\ref{tab:precision-by-score}).
This validates the rubric: the 1--10 score carries genuine
information about label confidence and is not merely an artifact of
the prompt template. Practitioners requiring higher precision than
the corpus-level 96.98\% can simply raise the score threshold; the
released corpus uses the inclusive $\geq 8$ cut to maximize coverage.

\begin{table}[ht]
\centering
\caption{Precision conditional on the GPT-4.1 score, gold sample
($n=2{,}421$).}
\label{tab:precision-by-score}
\begin{tabular}{rrrrl}
\toprule
GPT score & $n$ & $n_{\text{correct}}$ & Precision & 95\% CI \\
\midrule
8 & 314 & 285 & 0.9076 & [0.8705, 0.9349] \\
9 & 1{,}965 & 1{,}922 & 0.9781 & [0.9707, 0.9837] \\
10 & 142 & 141 & 0.9930 & [0.9612, 0.9988] \\
\bottomrule
\end{tabular}
\end{table}

\subsection{Stratified error analysis}
The error rate is concentrated in two sectors. In Manufacturing
(31--33), repositories at score 8 are correct only $20\%$ of the time
(4 of 20 gold rows), but rise to $80.5\%$ at score 9 and $100\%$ at
score 10. Wholesale Trade (42) shows the same pattern. Practitioners
building on NAICS-GH who require high precision in these two sectors
should raise the score threshold to $\geq 9$. A qualitative analysis
of the 73 incorrect labels --- which NAICS sector RAs would have
assigned instead of the model's prediction --- is in
Appendix~\ref{app:error-analysis}.

\subsection{What this validation does \emph{not} cover}
\label{sec:val:notcover}

\paragraph{Single-annotator protocol.} Each gold-set row was reviewed
by exactly one research assistant, so we cannot report Cohen's
$\kappa$ or another inter-annotator agreement metric. A future
double-labeling pass on $\sim 200$ rows will let an agreement
coefficient accompany the headline precision.

\paragraph{Precision, not recall.} The validation tells us what
fraction of \emph{retained} labels are correct, but not what fraction
of true sector-$X$ repositories in the source pool the pipeline
\emph{missed} during retrieval. A recall analysis would require an
independently constructed reference list of sector-labeled
repositories from a source other than the BGE/FAISS retrieval used
here, and is left to future work.

\section{Benchmark: Fine-Tuned NAICS Classifiers}
\label{sec:benchmark}

To demonstrate the downstream utility of NAICS-GH, we fine-tune six
pretrained encoders on the released 6{,}588-row training corpus and
compare them on a held-out test set. All six runs share identical
data, splits, and hyperparameters --- only the base model differs.

\subsection{Input construction}

Each repository is serialized as

\begin{center}
\small
\texttt{Repository: \{name\_repo\} | Description: \{description\} | Topics: \{topics\} | README: \{readme\_content\}}
\end{center}

\noindent and then passed through a cleaner that strips badges, license
headers, Markdown formatting, code blocks (which are replaced by a
\texttt{code-\{lang\}} placeholder), excess punctuation, installation
commands (\texttt{npm install}, \texttt{pip install},
\texttt{git clone}), and collapses URLs to their domain. The cleaned
string is truncated by the tokenizer to a uniform \textbf{512
WordPiece tokens} across all six models (we choose this max length for
direct comparability, even for the ModernBERT variants that support
longer contexts).

\subsection{Splits and hyperparameters}

We use a stratified 70 / 10 / 20 train / validation / test split with
\texttt{random\_state = 42}, yielding sizes
\textbf{4{,}611 / 659 / 1{,}318}. All 19 NAICS classes are present in
every split. Training uses AdamW
(\texttt{adamw\_torch\_fused}) with learning rate $1.5\times 10^{-5}$,
polynomial schedule with $15\%$ warmup, weight decay $0.02$,
per-device batch size 8, gradient-accumulation steps 2 (effective
batch size 16), gradient clipping at 1.0, BF16 mixed precision, 8
epochs, and early stopping with patience 2 and threshold $0.001$ on
weighted F1. Evaluation and checkpoint saving fire every 100 steps,
with the best checkpoint by F1 loaded at the end of training. The
reported metrics are weighted F1, accuracy, precision, and recall on
the held-out test set.

\subsection{Models and results}

Table~\ref{tab:bench} reports test-set performance for six pretrained
encoders spanning three families
(RoBERTa~\citep{liu2019roberta},
ModernBERT~\citep{warner2024modernbert},
DeBERTa-v3~\citep{he2023debertav3}) and two
parameter scales (base and large).

\begin{table}[ht]
\centering
\caption{Test-set performance on the 1,318-row held-out split, all
metrics weighted across the 19 NAICS classes.}
\label{tab:bench}
\small
\begin{tabular}{lrrrrr}
\toprule
Model & Parameters & Test F1 & Accuracy & Precision & Recall \\
\midrule
RoBERTa-base & 125M & 84.26\% & 84.84\% & 84.70\% & 84.84\% \\
\textbf{RoBERTa-large} & \textbf{355M} & \textbf{86.45\%} & \textbf{86.35\%} & \textbf{86.68\%} & \textbf{86.35\%} \\
ModernBERT-base & 139M & 84.63\% & 84.85\% & 84.86\% & 84.85\% \\
ModernBERT-large & 395M & 84.16\% & 84.23\% & 84.44\% & 84.23\% \\
DeBERTa-v3-base & 183M & 85.68\% & 85.82\% & 86.16\% & 85.82\% \\
DeBERTa-v3-large & 400M & 85.07\% & 85.22\% & 85.41\% & 85.22\% \\
\bottomrule
\end{tabular}
\end{table}

\paragraph{Findings.}
\begin{itemize}
  \item \textbf{RoBERTa-large} is the strongest baseline at
    F1 $=86.45\%$, followed by DeBERTa-v3-base ($85.68\%$) and
    DeBERTa-v3-large ($85.07\%$).
  \item \textbf{The encoder family matters more than parameter
    count} at this corpus size. DeBERTa-v3-base (183M) outperforms
    ModernBERT-large (395M), suggesting that the original
    RoBERTa/DeBERTa pretraining objectives transfer better to
    short-document classification than ModernBERT's
    long-context-oriented pretraining, at least for a corpus on the
    order of $5\,000$ training examples.
  \item \textbf{Larger variants do not uniformly outperform their
    base counterparts.} ModernBERT-large is worse than
    ModernBERT-base, and DeBERTa-v3-large is worse than
    DeBERTa-v3-base. RoBERTa is the only family where scaling
    monotonically improves performance, which is consistent with a
    relatively small training corpus (4.6k examples, 19 classes)
    where larger models risk overfitting.
\end{itemize}

The fine-tuned RoBERTa-large checkpoint is available at
\url{https://huggingface.co/alexanderquispe/naics-github-classifier}.

\section{Limitations and Ethical Considerations}
\label{sec:limitations}

\paragraph{English-only retrieval.} BGE-large-en is trained on English
text. Repositories whose READMEs are in other languages are
under-represented in the candidate pool and therefore in the released
corpus. We make no claim of coverage for non-English software ecosystems.

\paragraph{NAICS is a North American taxonomy.} Applying NAICS to
European or Australian repositories assumes that economic activities
map cleanly across jurisdictions. Sector definitions sometimes diverge
(for example, NAICS Sector 22 ``Utilities'' is structured around the
US regulated-utility model). We note this whenever it materially affects
interpretation.

\paragraph{Label noise is not uniform.} As Section~\ref{sec:validation}
shows, two sectors (Manufacturing and Wholesale Trade) carry
substantially higher label error than the rest at the score-$\geq 8$
threshold. Downstream users should treat the GPT score as a usable
confidence signal (raise to $\geq 9$ for stricter applications)
rather than treat all labels as a uniform gold standard.

\paragraph{Repository content licensing.} NAICS-GH releases labels
\emph{about} public repositories, plus excerpts of their READMEs for
reproducibility. The underlying repositories remain governed by their
own licenses (preserved as \texttt{spdx\_license} in the released file).

\paragraph{Dual use.} Industry-classified repository data could
plausibly be used for competitive intelligence or surveillance of
open-source contributors. We release the dataset under CC-BY-4.0 with a
preferred-use statement encouraging academic and policy use; we do not
believe sector-level industry tagging of public repositories raises
additional risks beyond what the underlying public metadata already
permits.

\section{Release and Reproducibility}
\label{sec:release}

The dataset is released at
\url{https://huggingface.co/datasets/aquiro1994/naics-gh} under
CC-BY-4.0, with stratified train/validation/test splits
($4{,}611 / 659 / 1{,}318$) ready for direct use via
\texttt{datasets.load\_dataset("aquiro1994/naics-gh")}. Croissant
metadata is generated automatically by Hugging Face from the dataset
card's YAML front matter, satisfying the NeurIPS Datasets \&
Benchmarks 2025 machine-readable-metadata requirement. The labeling
pipeline code is at
\url{https://github.com/alexanderquispe/naics-github-classifier} (MIT
License); the training code and fine-tuned RoBERTa-large checkpoint
are at
\url{https://github.com/alexanderquispe/naics-github-train} and
\url{https://huggingface.co/alexanderquispe/naics-github-classifier}.
A Zenodo DOI for the dataset will be issued at camera-ready.

\paragraph{End-to-end replication of the labeling pipeline.}
To quantify the reproducibility of the dataset-generation process, we
re-ran the full retrieval pipeline from the frozen source extractions
on different hardware (consumer RTX~3080 vs.\ the original A100). The
raw retrieval depth is exactly reproduced by construction
(20{,}879 rows per region), and the deduplicated candidate sets match
to within $\pm 0.1\%$ per region --- 10{,}985 vs.\ 10{,}983 (USA),
10{,}529 vs.\ 10{,}531 (EU), and 9{,}674 vs.\ 9{,}664 (AU); 31{,}188
vs.\ 31{,}178 in total ($+0.03\%$). The residual differences are
attributable to FP16 floating-point variation across GPU
architectures shifting a handful of borderline nearest neighbors.

\paragraph{Jurisdiction-blind public replication package.}
The released corpus was generated by the per-region pipeline described
in \S\ref{sec:source:merge}, which requires access to
jurisdiction-tagged source extractions that we do not publish. To let
external researchers reproduce the \emph{methodology} end-to-end, the
pipeline repository additionally provides a jurisdiction-blind variant
that concatenates the three extractions into a single shuffled pool
before embedding and retrieves with a proportionally scaled depth
(base $k$ tripled). Because a merged pool deduplicates the
cross-region overlap that separate regional pools structurally retain,
this variant yields a corpus of comparable --- not bit-identical ---
size and composition; it exists to verify the method, while the
released corpus remains the canonical artifact.

\section{Conclusion}
\label{sec:conclusion}

NAICS-GH supplies a missing resource: a publicly available mapping
from GitHub repositories to standardized industry sectors,
GPT-4.1-labeled over a multi-jurisdictional source pool and validated
against a human-labeled random subsample at 96.98\% precision. We
hope it enables empirical work on the industrial composition of
open-source production --- over time and in response to the rise of
AI coding assistants. An inter-annotator-agreement coefficient on a
doubly labeled subsample and a recall estimate against an independent
reference list of sector-labeled repositories are concrete next
steps.

\bibliographystyle{plainnat}
\bibliography{references}

\appendix

\section{Source-extraction SQL}
\label{app:sql}

The canonical Presto/Trino query against GitHub's internal warehouse
that produced the source parquets. The query is executed three times,
varying only the \texttt{accounts\_current.country\_account IN (\dots)}
clause (see \S\ref{sec:source:regions}); each output is processed
per region by an identical pipeline configuration
(\S\ref{sec:source:merge}), and the \texttt{country\_code} column is
dropped at that
point.

{\small
\begin{verbatim}
SELECT
    nwo, num_stars, is_fork, is_archived, has_readme,
    content_size AS readme_content_size, disk_usage_bytes,
    spdx_license,
    repositories_current.id AS repository_id,
    from_utf8(repositories_current.description) AS description,
    repositories_current.topics,
    from_utf8(readme_current.content) AS readme_content,
    accounts_current.country_account AS country_code,
    COUNT(DISTINCT issues_current.id) AS num_issues,
    COUNT(DISTINCT issues_current.user_dotcom_id) AS num_issue_authors,
    COUNT(DISTINCT github_collab_commit_contributions.user_id)
        AS num_commit_contributors,
    SUM(github_collab_commit_contributions.commit_count) AS num_commits,
    MAX(github_collab_commit_contributions.created_at) AS last_commit_at
FROM hive.canonical.repositories_current
JOIN hive.canonical.accounts_current
    ON repositories_current.owner_dotcom_id = accounts_current.dotcom_id
LEFT JOIN hive.canonical.issues_current
    ON issues_current.repository_id = repositories_current.id
    AND NOT issues_current.is_user_hidden
JOIN hive.suez.readme_current
    ON readme_current.repository_id = repositories_current.id
    AND LOWER(path) = 'readme.md'
    AND LOWER(filename) = 'readme.md'
LEFT JOIN hive.reference.licenses
    ON licenses.id = repositories_current.license_id
LEFT JOIN delta.snapshots.github_collab_commit_contributions
    ON github_collab_commit_contributions.repository_id
       = repositories_current.id
WHERE NOT is_spammy_owner
    AND is_public = TRUE
    AND has_readme = TRUE
    AND num_stars >= 1
    AND is_fork = FALSE
    AND accounts_current.country_account IN ( /* per-region list */ )
    AND content_size >= 750
    AND disk_usage_bytes > 0
    AND repositories_current.description IS NOT NULL
    AND TRIM(from_utf8(repositories_current.description)) != ''
GROUP BY 1,2,3,4,5,6,7,8,9,10,11,12
HAVING SUM(github_collab_commit_contributions.commit_count) > 5
   AND COUNT(DISTINCT github_collab_commit_contributions.user_id) >= 2
ORDER BY repository_id
\end{verbatim}
}

The warehouse data was current through April 15, 2025 (USA presample);
the EU README extraction snapshot was dated August 15, 2025.

\section{NAICS sector definitions}
\label{app:naics}

We use the 20-sector top-level NAICS 2022 hierarchy as compiled by
the U.S. Census Bureau~\citep{naics2022}. The pipeline consumes a
flat JSON taxonomy file
(\texttt{naics\_titles\_by\_group\_6digit\_clean.json}, released with
the code) keyed by 2-digit sector code; each value is a
semicolon-separated string of 6-digit subindustry titles drawn from
the 2022 hierarchy. Across the 20 sectors there are 1{,}029 distinct
subindustry phrases.

Table~\ref{tab:naics-appendix} reports, for each 2-digit sector, the
canonical Census name, the number of subindustry phrases in our
taxonomy file, and the first three subindustry phrases as they
appear in the JSON (verbatim, including any minor whitespace
artifacts). The full per-sector listings are in the JSON file rather
than this appendix in order to keep the paper a reasonable length.

\input{appendix/output/naics_appendix_table.tex}

\section{LLM verification prompt}
\label{app:prompt}

\input{appendix/output/prompt_appendix.tex}

\section{Schema of the released file}
\label{app:schema}

\input{appendix/output/schema_appendix.tex}

\section{Stratification of the gold sample}
\label{app:stratification}

\input{appendix/output/stratification_appendix.tex}

\section{Training details}
\label{app:training}

\input{appendix/output/training_appendix.tex}

\section{Qualitative error analysis}
\label{app:error-analysis}

\section{Datasheet for Datasets}
\label{app:datasheet}

\input{datasheet_body.tex}

\end{document}

%% file: validation/output/sector_counts_table.tex
\resizebox{\linewidth}{!}{%
\begin{tabular}{llr}
\toprule
Code & Sector & $n$ \\
\midrule
11 & Agriculture, Forestry, Fishing and Hunting & 466 \\
21 & Mining & 86 \\
22 & Utilities & 414 \\
23 & Construction & 82 \\
31-33 & Manufacturing & 311 \\
42 & Wholesale Trade & 121 \\
44-45 & Retail Trade & 641 \\
48-49 & Transportation and Warehousing & 558 \\
51 & Information & 389 \\
52 & Finance and Insurance & 429 \\
53 & Real Estate Rental and Leasing & 284 \\
54 & Professional and Technical Services & 372 \\
56 & Administrative and Support Services & 184 \\
61 & Educational Services & 331 \\
62 & Health Care and Social Assistance & 466 \\
71 & Arts, Entertainment, and Recreation & 453 \\
72 & Accomodation and Food Services & 333 \\
81 & Other Services & 295 \\
92 & Public Administration & 373 \\
\midrule
\multicolumn{2}{l}{\textbf{Total}} & \textbf{6,588} \\
\bottomrule
\end{tabular}%
}

%% file: validation/output/per_sector_precision_table.tex
\resizebox{\linewidth}{!}{%
\begin{tabular}{llrrl}
\toprule
Code & Sector & $n$ & Precision & 95\% CI \\
\midrule
92 & Public Administration & 184 & 1.000 & [0.980, 1.000] \\
51 & Information & 148 & 1.000 & [0.975, 1.000] \\
71 & Arts, Entertainment, and Recreation & 176 & 1.000 & [0.979, 1.000] \\
53 & Real Estate Rental and Leasing & 91 & 1.000 & [0.959, 1.000] \\
21 & Mining & 32 & 1.000 & [0.893, 1.000] \\
48-49 & Transportation and Warehousing & 200 & 1.000 & [0.981, 1.000] \\
72 & Accomodation and Food Services & 117 & 0.991 & [0.953, 0.998] \\
61 & Educational Services & 115 & 0.991 & [0.952, 0.998] \\
52 & Finance and Insurance & 166 & 0.988 & [0.957, 0.997] \\
81 & Other Services & 116 & 0.983 & [0.939, 0.995] \\
62 & Health Care and Social Assistance & 204 & 0.980 & [0.951, 0.992] \\
44-45 & Retail Trade & 202 & 0.980 & [0.950, 0.992] \\
22 & Utilities & 157 & 0.975 & [0.936, 0.990] \\
54 & Professional and Technical Services & 115 & 0.974 & [0.926, 0.991] \\
56 & Administrative and Support Services & 62 & 0.968 & [0.890, 0.991] \\
23 & Construction & 25 & 0.960 & [0.805, 0.993] \\
11 & Agriculture, Forestry, Fishing and Hunting & 174 & 0.931 & [0.883, 0.960] \\
31-33 & Manufacturing & 100 & 0.730 & [0.636, 0.807] \\
42 & Wholesale Trade & 37 & 0.730 & [0.570, 0.846] \\
\bottomrule
\end{tabular}%
}

%% file: appendix/output/naics_appendix_table.tex
\begin{longtable}{@{}lp{4.3cm}rp{6.0cm}@{}}
\caption{NAICS 2022 sectors used by the pipeline. The $n$ column gives the number of 6-digit subindustry phrases under each 2-digit sector in our taxonomy file; the full list of 1{,}029 phrases is in \texttt{naics\_titles\_by\_group\_6digit\_clean.json} in the released code repository. The Example column shows the first 3 subindustry phrases as they appear in the file (verbatim, including any minor whitespace). Sector~55 is absent from the released training corpus because too few candidates passed verification under the 80-sample minimum-class-size filter; it is retained here for taxonomic completeness.}\\
\label{tab:naics-appendix}\\
\toprule
Code & Sector & $n$ & Example subindustry titles \\
\midrule
\endfirsthead
\multicolumn{4}{c}{\textit{(continued from previous page)}}\\
\toprule
Code & Sector & $n$ & Example subindustry titles \\
\midrule
\endhead
\bottomrule
\endlastfoot
11 & Agriculture, Forestry, Fishing and Hunting & 64 & Soybean Farming; Oilseed Farming; Dry Pea and Bean Farming; \dots \\
21 & Mining, Quarrying, and Oil and Gas Extraction & 21 & Crude Petroleum Extraction; Natural Gas Extraction; Surface Coal Mining; \dots \\
22 & Utilities & 14 & Hydroelectric Power Generation; Fossil Fuel Electric Power Generation; Nuclear Electric Power Generation; \dots \\
23 & Construction & 31 & New Single-Family Housing Construction; New Multifamily Housing Construction; New Housing For-Sale Builders; \dots \\
31-33 & Manufacturing & 346 & Dog and Cat Food Manufacturing; Other Animal Food Manufacturing; Flour Milling; \dots \\
42 & Wholesale Trade & 69 & Automobile and Other Motor Vehicle Merchant Wholesalers; Motor Vehicle Supplies and New Parts Merchant Wholesalers; Tire and Tube Merchant Wholesalers; \dots \\
44-45 & Retail Trade & 57 & New Car Dealers; Used Car Dealers; Recreational Vehicle Dealers; \dots \\
48-49 & Transportation and Warehousing & 57 & Scheduled Passenger Air Transportation; Scheduled Freight Air Transportation; Nonscheduled Chartered Passenger Air Transportation; \dots \\
51 & Information & 29 & Motion Picture and Video Production; Motion Picture and Video Distribution; Motion Picture Theaters; \dots \\
52 & Finance and Insurance & 35 & Monetary Authorities-Central Bank; Commercial Banking; Credit Unions; \dots \\
53 & Real Estate and Rental and Leasing & 24 & Lessors of Residential Buildings and Dwellings; Lessors of Nonresidential Buildings; Lessors of Miniwarehouses and Self-Storage Units; \dots \\
54 & Professional, Scientific, and Technical Services & 49 & Offices of Lawyers; Offices of Notaries; Title Abstract and Settlement Offices; \dots \\
55 & Management of Companies and Enterprises & 20 & Bank holding companies; Holding companies, bank; Offices of bank holding companies; \dots \\
56 & Administrative and Support and Waste Management and Remediation Services & 44 & Office Administrative Services; Facilities Support Services; Employment Placement Agencies; \dots \\
61 & Educational Services & 17 & Elementary and Secondary Schools; Junior Colleges; Colleges, Universities, and Professional Schools; \dots \\
62 & Health Care and Social Assistance & 39 & Offices of Physicians; Offices of Physicians, Mental Health Specialists; Offices of Dentists; \dots \\
71 & Arts, Entertainment, and Recreation & 25 & Theater Companies and Dinner Theaters; Dance Companies; Musical Groups and Artists; \dots \\
72 & Accommodation and Food Services & 15 & Hotels and Motels; Casino Hotels; Bed-and-Breakfast Inns; \dots \\
81 & Other Services (except Public Administration) & 44 & General Automotive Repair; Specialized Automotive Repair; Automotive Body, Paint, and Interior Repair and Maintenance; \dots \\
92 & Public Administration & 29 & Executive Offices; Legislative Bodies; Public Finance Activities; \dots \\
\end{longtable}

%% file: appendix/output/prompt_appendix.tex
This appendix reproduces verbatim the prompt sent to GPT-4.1 for every candidate (repository, NAICS sector) pair. The three placeholders \texttt{\{naics\_code\}}, \texttt{\{readme\_text.strip()\}}, and \texttt{\{sector\_industries.strip()\}} are substituted per row before the request is sent. Source: \texttt{3\_eu\_repo\_gpt\_classif.ipynb} cell 22 (\texttt{build\_prompt} function) and cell 26 (system message and API call). The corresponding Python mirror is \texttt{3\_eu\_repo\_gpt\_classif.py}.

\paragraph{System message.}
\begin{Verbatim}[fontsize=\footnotesize,breaklines=true,breakanywhere=true,breakindent=0pt,breaksymbolleft={\textcolor{gray}{\tiny$\hookrightarrow$}},frame=single,framesep=2mm,xleftmargin=0pt,xrightmargin=0pt]
You are a domain expert in economic classification systems with a focus on NAICS industry {naics_code}.
\end{Verbatim}

\paragraph{User message.}
\begin{Verbatim}[fontsize=\footnotesize,breaklines=true,breakanywhere=true,breakindent=0pt,breaksymbolleft={\textcolor{gray}{\tiny$\hookrightarrow$}},frame=single,framesep=2mm,xleftmargin=0pt,xrightmargin=0pt]
TASK: GitHub Repository NAICS Sector Classification

You are a domain expert tasked with classifying GitHub repositories into NAICS industry sectors for GitHub's repository categorization system. Your goal is to determine with high precision whether this repository belongs to NAICS Sector {naics_code}.

REPOSITORY README CONTENT:
<readme>
{readme_text.strip()}
</readme>

TARGET NAICS SECTOR:
<naics_sector>
Sector {naics_code}: {sector_industries.strip()}
</naics_sector>

CLASSIFICATION FRAMEWORK:

A repository should be classified as "Yes" for this sector if it demonstrates CLEAR ALIGNMENT with one or more of these criteria:

1. **Industry-Specific Software**: Applications, tools, or systems designed specifically for use within this industry sector
   - Example: Farm management software for Agriculture (Sector 11)
   - Example: Church management systems for Religious Organizations (Sector 81)

2. **Sector-Relevant Functionality**: Code that implements processes, calculations, or workflows specific to this industry
   - Example: Prayer time calculators for Religious Organizations
   - Example: Crop yield prediction models for Agriculture

3. **Industry Domain Applications**: Software that directly serves businesses, organizations, or activities within this sector
   - Example: Restaurant POS systems for Food Services (Sector 72)
   - Example: Educational platforms for Educational Services (Sector 61)

4. **Sector-Specific Data/Research**: Datasets, analysis tools, or research implementations focused on this industry
   - Example: Agricultural sensor data analysis
   - Example: Healthcare outcome prediction models

CLASSIFICATION STANDARDS:

**INCLUDE ("Yes") when:**
- The repository's primary purpose aligns with the sector
- The software would be used by businesses/organizations in this sector
- The code implements sector-specific functionality or processes
- The project addresses sector-specific problems or use cases

**EXCLUDE ("No") when:**
- The repository serves multiple sectors equally (generic tools)
- Industry connection is only tangential or in examples
- The primary use case is outside this sector
- No clear business or operational relevance to the sector

SCORING GUIDE:
- 9-10: Core industry software with direct sector application
- 7-8: Strong sector relevance with clear industry use cases
- 5-6: Moderate sector connection with identifiable applications
- 3-4: Weak sector relevance, mostly tangential
- 1-2: No meaningful sector connection

ANALYSIS REQUIREMENTS:
1. Identify the repository's primary purpose and functionality
2. Assess alignment with the target NAICS sector
3. Determine the most applicable classification criterion
4. Consider practical usage within the sector

Provide your response in this exact JSON format:
{
    "NAICS {naics_code}": {
        "rationale": "Concise explanation of classification decision, including primary repository purpose, specific sector alignment criteria met, and justification for inclusion/exclusion",
        "score": "1-10",
        "match": "Yes" or "No"
    }
}

IMPORTANT: Base your decision on the repository's PRIMARY purpose and DIRECT applicability to the sector. Be precise and consistent in your classifications.
\end{Verbatim}

\paragraph{API parameters.}
\begin{itemize}
  \item \textbf{Endpoint:} \url{https://api-model-lab.githubcopilot.com/chat/completions}
  \item \textbf{Model alias:} \texttt{gpt-4.1} (resolves to snapshot \texttt{gpt-4.1-2025-04-14} at the GitHub Copilot LLM-lab endpoint).
  \item \textbf{\texttt{temperature}:} 0
  \item \textbf{\texttt{max\_tokens} (output):} $\min(3500,\; 128000 - n_{\text{input}})$, where $n_{\text{input}}$ is the token count of the system plus user message under \texttt{tiktoken.encoding\_for\_model(\"gpt-4.1\")}.
  \item \textbf{Retry policy:} up to 5 attempts on any non-200 response, with linear back-off \texttt{sleep = 5 $\times$ attempt} seconds.
  \item \textbf{Response parsing:} the model's reply is searched with \texttt{re.search(r'\textbackslash\{[\textbackslash s\textbackslash S]*\textbackslash\}', reply)} and the matched substring passed to \texttt{json.loads}. The expected output is nested under the outer key \texttt{\"NAICS \{code\}\"} with string-valued \texttt{match} (\texttt{\"Yes\"} or \texttt{\"No\"}), string-valued \texttt{score} (\texttt{\"1\"}--\texttt{\"10\"}), and free-form \texttt{rationale}.
\end{itemize}

%% file: appendix/output/schema_appendix.tex
The released file \texttt{train\_data\_gpt\_ab8\_score\_with\_code.parquet} contains \textbf{6,588 rows} and \textbf{6 columns}. Every cell is non-null. Table~\ref{tab:schema-fields} gives the per-column schema; Table~\ref{tab:schema-stats} gives character-length and uniqueness statistics for the four text columns; the listing at the end shows one full row (with the \texttt{readme\_content} field truncated for readability).

\begin{table}[ht]
\centering
\caption{Schema of the released training corpus. The order shown matches the column order in the parquet file.}
\label{tab:schema-fields}
\small
\begin{tabular}{lll}
\toprule
Column & Type & Description \\
\midrule
\texttt{name\_repo} & string & Repository short name (no owner prefix). \\
\texttt{description} & string & Repository description from GitHub. \\
\texttt{topics} & string & Topic tags joined with ``\texttt{;\space}'', empty string if the repository declares no tags. \\
\texttt{readme\_content} & string & Cleaned README text (Markdown code blocks, HTML tags, and URLs stripped; whitespace normalized). \\
\texttt{label} & int64 & Integer class encoding, $0\ldots 18$, monotonic in \texttt{code}. \\
\texttt{code} & string & 2-digit NAICS sector code as a string. \\
\bottomrule
\end{tabular}
\end{table}

\begin{table}[ht]
\centering
\caption{Per-column character-length and uniqueness statistics for the four string columns. Empty strings (length 0) count as present, not missing.}
\label{tab:schema-stats}
\small
\begin{tabular}{lrrrrrr}
\toprule
Column & Min & Median & Mean & Max & Empty rows & Unique \\
\midrule
\texttt{name\_repo} & 2 & 14 & 16 & 85 & 0 (0.0\%) & 6,141 \\
\texttt{description} & 2 & 65 & 88 & 1,626 & 0 (0.0\%) & 6,231 \\
\texttt{topics} & 0 & 0 & 21 & 384 & 4,533 (68.8\%) & 1,926 \\
\texttt{readme\_content} & 3 & 1912 & 3387 & 212,122 & 0 (0.0\%) & 6,192 \\
\bottomrule
\end{tabular}
\end{table}

\paragraph{Example row.}
One representative row from the released file (the \texttt{readme\_content} field is truncated here to 240 characters for readability; the actual stored value may be much longer).

\begin{Verbatim}[fontsize=\footnotesize,breaklines=true,breakanywhere=true,breakindent=0pt,breaksymbolleft={\textcolor{gray}{\tiny$\hookrightarrow$}},frame=single,framesep=2mm,xleftmargin=0pt,xrightmargin=0pt]
name_repo:      etn-occurrences
description:    Acoustic telemetry data
topics:         lifewatch; oscibio; animal-tracking; data-publication; dataset; biologging; animal-movement; rstats; fish; r
readme_content: # Acoustic telemetry datasets This repository contains scripts to publish fish tracking data from the [European Tracking Network (ETN)](lifewatch.be (specifically from the [Permanent Belgian Acoustic Receiver Network](lifewatch.be on [GBIF...
label:          0
code:           11
\end{Verbatim}

%% file: appendix/output/stratification_appendix.tex
This appendix tabulates the per-sector distribution of the validation gold sample (\S\ref{sec:validation}; $n = 2,421$ rows) against the released NAICS-GH training corpus ($n = 6,588$ rows). The largest signed delta is $\pm 1.94$ percentage points; 16 of 19 sectors fall within $\pm 1$ pp.

Reasonable agreement supports our claim that the headline 96.98\% precision is not biased by an over- or under-representation of any single NAICS sector in the gold sample.

\begin{table}[ht]
\centering
\caption{Per-sector distribution: validation gold sample vs.\ the released NAICS-GH corpus. \emph{Gold \%} is the share of the 2,421 gold rows in each sector; \emph{Corpus \%} is the share of the 6,588 released rows; \emph{$\Delta$} is the signed difference in percentage points (gold $-$ corpus).}
\label{tab:stratification}
\small
\begin{tabular}{llrrrrr}
\toprule
Code & Sector & Gold $n$ & Corpus $n$ & Gold \% & Corpus \% & $\Delta$ (pp) \\
\midrule
11 & Agriculture, Forestry, Fishing and Hunting & 174 & 466 & 7.19 & 7.07 & +0.11 \\
21 & Mining & 32 & 86 & 1.32 & 1.31 & +0.02 \\
22 & Utilities & 157 & 414 & 6.48 & 6.28 & +0.20 \\
23 & Construction & 25 & 82 & 1.03 & 1.24 & -0.21 \\
31-33 & Manufacturing & 100 & 311 & 4.13 & 4.72 & -0.59 \\
42 & Wholesale Trade & 37 & 121 & 1.53 & 1.84 & -0.31 \\
44-45 & Retail Trade & 202 & 641 & 8.34 & 9.73 & -1.39 \\
48-49 & Transportation and Warehousing & 200 & 558 & 8.26 & 8.47 & -0.21 \\
51 & Information & 148 & 389 & 6.11 & 5.90 & +0.21 \\
52 & Finance and Insurance & 166 & 429 & 6.86 & 6.51 & +0.34 \\
53 & Real Estate Rental and Leasing & 91 & 284 & 3.76 & 4.31 & -0.55 \\
54 & Professional and Technical Services & 115 & 372 & 4.75 & 5.65 & -0.90 \\
56 & Administrative and Support Services & 62 & 184 & 2.56 & 2.79 & -0.23 \\
61 & Educational Services & 115 & 331 & 4.75 & 5.02 & -0.27 \\
62 & Health Care and Social Assistance & 204 & 466 & 8.43 & 7.07 & +1.35 \\
71 & Arts, Entertainment, and Recreation & 176 & 453 & 7.27 & 6.88 & +0.39 \\
72 & Accomodation and Food Services & 117 & 333 & 4.83 & 5.05 & -0.22 \\
81 & Other Services & 116 & 295 & 4.79 & 4.48 & +0.31 \\
92 & Public Administration & 184 & 373 & 7.60 & 5.66 & +1.94 \\
\midrule
\textbf{Total} & & \textbf{2,421} & \textbf{6,588} & \textbf{100.00} & \textbf{100.00} & \textbf{0.00} \\
\bottomrule
\end{tabular}
\end{table}

%% file: appendix/output/training_appendix.tex
This appendix documents the full benchmark setup behind Section~\ref{sec:benchmark}: the input cleaning, the six base models and their parameter counts, the complete list of training hyperparameters (identical across all six runs), and reproducibility notes. The canonical source is \texttt{naics\_training\_gptdata.ipynb} in the Colab \texttt{naics\_github} folder; a re-runnable local refactor is \texttt{naics-github-train/scripts/train.py}.

\subsection*{F.1\quad Input construction and cleaning}
Each row is serialized as \texttt{Repository: \{name\_repo\} | Description: \{description\} | Topics: \{topics\} | README: \{readme\_content\}} and then passed through the helper \texttt{clean\_repository\_text(text)} which performs nine normalization steps:
\begin{enumerate}
  \item Strip Markdown badges and shields (\texttt{!{[}\dots{]}({[}url{]})} and \texttt{{[}!{[}\dots{]}(\dots){]}(\dots)});
  \item strip license/copyright headers (\texttt{MIT License}, \texttt{Apache License}, \texttt{GPL}, \texttt{BSD}, \texttt{Copyright \dots});
  \item collapse URLs to the bare domain (\texttt{https?://(domain)/path?...} $\rightarrow$ \texttt{domain});
  \item strip Markdown headers (\texttt{\#}\dots\texttt{\#\#\#\#\#\#}) and bold/italic/code markers (\texttt{*}, \texttt{\_}, \texttt{\textasciitilde}, \texttt{\textasciigrave});
  \item replace fenced code blocks with a \texttt{code-\{lang\}} placeholder; strip inline backticks but keep the content;
  \item normalize tech-stack mentions (\texttt{js} $\rightarrow$ \texttt{javascript}, \texttt{py} $\rightarrow$ \texttt{python}, \texttt{reactjs} $\rightarrow$ \texttt{react}, \texttt{nodejs} $\rightarrow$ \texttt{nodejs});
  \item normalize excessive punctuation (\texttt{!!} $\rightarrow$ \texttt{!}, \texttt{??} $\rightarrow$ \texttt{?}, \texttt{....} $\rightarrow$ \texttt{\dots});
  \item normalize whitespace (collapse multiple newlines and spaces);
  \item strip installation-command noise (\texttt{npm install}, \texttt{pip install}, \texttt{git clone}, and the rest of those lines).
\end{enumerate}
The cleaned string is the \texttt{text} column passed to the tokenizer.

\subsection*{F.2\quad Base models and parameter counts}
All six runs share data, splits, and hyperparameters; only the base model differs.
\begin{table}[ht]
\centering
\caption{Pretrained encoders fine-tuned in \S\ref{sec:benchmark}.}
\label{tab:training-models}
\small
\begin{tabular}{lll}
\toprule
Hugging Face identifier & Label in paper & Parameters \\
\midrule
\texttt{roberta-base} & RoBERTa-base & 125M \\
\texttt{roberta-large} & RoBERTa-large & 355M \\
\texttt{answerdotai/ModernBERT-base} & ModernBERT-base & 139M \\
\texttt{answerdotai/ModernBERT-large} & ModernBERT-large & 395M \\
\texttt{microsoft/deberta-v3-base} & DeBERTa-v3-base & 183M \\
\texttt{microsoft/deberta-v3-large} & DeBERTa-v3-large & 400M \\
\bottomrule
\end{tabular}
\end{table}

\subsection*{F.3\quad Training hyperparameters}
All values below are set in \texttt{setup\_training\_arguments()} and were applied identically to every model run.
\begin{table}[ht]
\centering
\caption{Full hyperparameter sheet for the fine-tuning runs.}
\label{tab:training-hp}
\small
\begin{tabular}{ll}
\toprule
Argument & Value \\
\midrule
\texttt{num\_train\_epochs} & 8 \\
\texttt{learning\_rate} & $1.5\times 10^{-5}$ \\
\texttt{lr\_scheduler\_type} & \texttt{polynomial} \\
\texttt{warmup\_ratio} & 0.15 \\
\texttt{weight\_decay} & 0.02 \\
\texttt{per\_device\_train\_batch\_size} & 8 \\
\texttt{per\_device\_eval\_batch\_size} & 16 \\
\texttt{gradient\_accumulation\_steps} & 2 \\
Effective batch size & 16 \\
\texttt{max\_grad\_norm} & 1.0 \\
\texttt{bf16} & \texttt{True} \\
\texttt{optim} & \texttt{adamw\_torch\_fused} \\
\texttt{eval\_strategy} & \texttt{steps} (every 100) \\
\texttt{save\_strategy} & \texttt{steps} (every 100) \\
\texttt{save\_total\_limit} & 5 \\
\texttt{load\_best\_model\_at\_end} & \texttt{True} \\
\texttt{metric\_for\_best\_model} & \texttt{f1} (weighted) \\
\texttt{early\_stopping\_patience} & 2 \\
\texttt{early\_stopping\_threshold} & 0.001 \\
\texttt{seed} & 42 \\
Tokenizer \texttt{max\_length} & 512 WordPiece tokens (uniform) \\
\bottomrule
\end{tabular}
\end{table}

\subsection*{F.4\quad Splits and evaluation metric}
The 6{,}588-row corpus is divided into train / validation / test sets of \textbf{4{,}611 / 659 / 1{,}318} via \texttt{sklearn.model\_selection.train\_test\_split} with \texttt{test\_size = 0.2}, \texttt{val\_size = 0.1}, \texttt{random\_state = 42}, and stratification on the \texttt{label} column. All 19 NAICS classes appear in every split. The evaluation metric for both \texttt{metric\_for\_best\_model} and the headline test-set result is the weighted F1 over the 19 classes; accuracy, weighted precision, and weighted recall are reported alongside.

\subsection*{F.5\quad Hardware and runtime}
All six runs were performed on a Colab Pro+ instance with a single NVIDIA A100 (40~GB). BF16 mixed precision and the fused \texttt{adamw\_torch\_fused} optimizer keep memory use comfortable for the largest model (DeBERTa-v3-large, 400~M parameters) at the per-device batch size of~8. End-to-end runtime per model was on the order of 8--15 minutes including checkpointing, with early stopping commonly firing between epochs 4 and 7.

\subsection*{F.6\quad Reproducibility}
The published RoBERTa-large checkpoint at \url{https://huggingface.co/alexanderquispe/naics-github-classifier} is the artifact produced by the above pipeline with \texttt{model\_id = \"roberta-large\"}. To reproduce, clone \texttt{naics-github-train}, place the released training file \texttt{train\_data\_gpt\_ab8\_score\_with\_code.parquet} under \texttt{data/raw/}, and run \texttt{python scripts/train.py --model roberta-large --batch-size 32 --epochs 8} (the script's defaults match the hyperparameters listed in Table~\ref{tab:training-hp}). With the same \texttt{seed = 42}, the F1 figure of 86.45\% should reproduce within $\pm 0.5$ pp of stochastic variation.

%% file: datasheet_body.tex
A ``Datasheet for Datasets'' following \href{https://dl.acm.org/doi/10.1145/3458723}{Gebru et al.~(2021)}.
This document accompanies the NAICS-GH dataset (Industry Classification of
GitHub Repositories Using the North American Industry Classification
System).

\begin{itemize}
\tightlist
\item
  \textbf{Dataset version:} v1.0
\item
  \textbf{Last updated:} 2026-05-22
\item
  \textbf{Authors:} Kevin Xu (GitHub), Alexander Quispe (GitHub)
\item
  \textbf{Contact:} alexanderquispe@github.com (corresponding)
\item
  \textbf{License:} CC-BY-4.0 (labels and metadata); MIT (pipeline code)
\end{itemize}

\begin{center}\rule{0.5\linewidth}{0.5pt}\end{center}

\hypertarget{motivation}{%
\subsection{1. Motivation}\label{motivation}}

\textbf{For what purpose was the dataset created?}
NAICS-GH was created to enable empirical work on the industrial
composition of open-source software production. GitHub hosts millions
of public repositories but provides no native indication of which
\emph{industry} a repository serves. NAICS-GH maps a representative subset
of repositories from the USA, the European Union, and Australia onto
the 2-digit North American Industry Classification System (NAICS),
which is the industry-classification standard used by US, Canadian, and
Mexican statistical agencies. The dataset is also the training data for
a downstream RoBERTa-large classifier that propagates these labels to
arbitrary repositories at inference time.

\textbf{Who created the dataset and on behalf of which entity?}
The dataset was created by Kevin Xu and Alexander Quispe at GitHub.

\textbf{Who funded the creation of the dataset?}
GitHub. Compute for LLM inference was paid through GitHub's internal
LLM access; no external grants were used.

\textbf{Any other comments?}
The dataset is one half of a two-repository system. This file
documents the labeled corpus; the downstream classifier and its
training code live in a sibling repository, \texttt{naics-github-train}.

\begin{center}\rule{0.5\linewidth}{0.5pt}\end{center}

\hypertarget{composition}{%
\subsection{2. Composition}\label{composition}}

\textbf{What do the instances represent?}
Each instance is a public GitHub repository, identified by its
\texttt{owner/name} (\texttt{nwo}) string. The labels attached to each instance
indicate which 2-digit NAICS industry sector the repository serves,
along with the LLM-generated rationale and confidence score.

\textbf{How many instances are there in total?}
The released corpus contains \textbf{6,588 repositories} in the parquet
file \texttt{train\_data\_gpt\_ab8\_score\_with\_code.parquet}. This is the file
that produces the published RoBERTa-large baseline.

\textbf{Does the dataset contain all possible instances, or is it a sample
from a larger set?}
It is a sample. The pipeline begins with three regional dumps of
public GitHub repositories (totaling 1,372,489 rows):
- USA: 510,380
- European Union: 530,898
- Australia: 331,211

For each NAICS subindustry (1,000+ across the 20 top-level sectors),
the top 20 most semantically similar repositories were retrieved using
BGE embeddings + FAISS, then LLM-scored. Only repositories scoring at
least 8 on the 1â€``10 rubric were retained.

\textbf{What data does each instance consist of?}
Each instance has six columns in the released file
(\texttt{train\_data\_gpt\_ab8\_score\_with\_code.parquet}):

\begin{longtable}[]{@{}lll@{}}
\toprule
Column & Type & Description \\
\midrule
\endhead
\texttt{name\_repo} & string & Repository short name (no owner prefix) \\
\texttt{description} & string & Repository description from GitHub \\
\texttt{topics} & string & Semicolon-joined topic tags; empty string if none \\
\texttt{readme\_content} & string & Cleaned README content \\
\texttt{label} & int64 & Integer class encoding 0â€``18 \\
\texttt{code} & string & 2-digit NAICS sector code \\
\bottomrule
\end{longtable}

The columns \texttt{nwo}, \texttt{match}, \texttt{score}, \texttt{rationale}, and \texttt{repo\_url}
present in intermediate files are intentionally dropped from the
public release so that the corpus is model-ready (no LLM-provenance,
no PII via repo-owner URLs).

\textbf{Is there a label or target associated with each instance?}
Yes. The label is the 2-digit NAICS sector code in the \texttt{code} column
(string form), or equivalently the integer \texttt{label} column (0â€``18,
monotonic in NAICS code order).

\textbf{Is any information missing from individual instances?}
- All four text columns and both label columns are fully populated
(zero NaN values).
- \texttt{topics} is empty for \textasciitilde69\% of rows (4,533 of 6,588): not all GitHub
repositories declare topic tags.
- \texttt{readme\_content} is preprocessed (code blocks stripped, HTML
removed, URLs collapsed) and truncated to 3,000 whitespace-separated
tokens before LLM scoring; the released field reflects the
\emph{preprocessed} text, not the original.

\textbf{Are there recommended data splits?}
For the RoBERTa-large baseline we used 70\% train / 10\% validation /
20\% test (\(n = 4{,}611 / 659 / 1{,}318\)) with \texttt{seed=42}. Splitting was
random and stratified by NAICS sector. Practitioners building their
own classifiers are free to re-split.

\textbf{Are there any errors, sources of noise, or redundancies?}
- \textbf{Label noise (USA scope only).} Human re-validation of 2,421
USA repositories found 96.98\% GPT-4.1 label precision overall, but
two sectors (31â€``33 Manufacturing and 42 Wholesale Trade) had only
\textasciitilde73\% precision at the score â‰¥ 8 threshold. EU and AU portions are
GPT-4.1-labeled by the same pipeline but have not yet been included
in a manual gold sample.
- \textbf{Duplicate \texttt{name\_repo}.} Repository short names are not globally
unique; the released file has 447 rows sharing a \texttt{name\_repo} with
another row, of which 113 share both \texttt{(name\_repo,\ code)}. These are
different repositories with identical short names that independently
passed verification. The upstream intermediate files retain \texttt{nwo}
as a globally unique identifier.
- \textbf{READMEs.} Some READMEs are predominantly HTML badges, build
configs, or templates that carry little semantic signal. The
preprocessing step removes most of these artifacts but cannot
recover content that wasn't there.

\textbf{Is the dataset self-contained, or does it link to or otherwise rely
on external resources?}
The released parquet is self-contained. The pipeline that generated it
relies on (a) BAAI/bge-large-en (released model, Hugging Face),
(b) FAISS (open source), and (c) GPT-4.1 (snapshot
\texttt{gpt-4.1-2025-04-14}) accessed via the GitHub Copilot LLM-lab
endpoint. The repository content (READMEs, descriptions) is preserved
in the file at the time of capture; we do not re-fetch live GitHub
content.

\textbf{Does the dataset contain data that might be considered confidential
or that includes content protected by attorney-client privilege or
similar?}
No.~All repositories are public on GitHub.

\textbf{Does the dataset contain data that, if viewed directly, might be
offensive, insulting, threatening, or might otherwise cause anxiety?}
Possibly, since READMEs contain free-form user-submitted text. We do
not apply explicit content filtering. Anyone re-using the README
content should be aware that public repository content can include
offensive language.

\textbf{Does the dataset identify any subpopulations (e.g., by age, gender)?}
No.~The dataset captures organizational and industrial-economic
attributes of repositories, not demographic attributes of contributors.

\textbf{Is it possible to identify individuals from the data?}
Repository owner names (\texttt{nwo} prefix) may identify individual GitHub
account holders, since public repositories often live under personal
accounts. This is the same level of identifiability already present
on GitHub itself; the dataset does not add information beyond what
GitHub publishes.

\textbf{Does the dataset contain data that might be considered sensitive in
any way?}
The dataset records that a given GitHub user's public repository
serves a given industry sector. We do not consider this sensitive
beyond the existing public nature of GitHub, but downstream uses
should respect the principle that aggregating labels about individuals
can amplify identifiability.

\begin{center}\rule{0.5\linewidth}{0.5pt}\end{center}

\hypertarget{collection-process}{%
\subsection{3. Collection Process}\label{collection-process}}

\textbf{How was the data associated with each instance acquired?}
Source repositories were extracted from GitHub's internal Trino-on-Hive
data warehouse via a Presto/Trino SQL query against the
\texttt{hive.canonical.repositories\_current}, \texttt{hive.canonical.accounts\_current},
\texttt{hive.suez.readme\_current}, and
\texttt{delta.snapshots.github\_collab\_commit\_contributions} tables. The
query yielded three regional dumps (USA: 510,380; EU: 530,898;
AU: 331,211 â€'' totaling 1,372,489 repositories). Each repository's
NAICS label was then \emph{derived} via a two-stage pipeline (see
Section 4). The label is not directly observed but is the output of
an algorithm that combines semantic retrieval with an LLM-scored
rubric.

\textbf{What mechanisms or procedures were used to collect the data?}
- The source SQL filters on \texttt{is\_public\ =\ TRUE}, \texttt{is\_fork\ =\ FALSE},
\texttt{NOT\ is\_spammy\_owner}, \texttt{num\_stars\ \textgreater{}=\ 1}, README size â‰¥ 750 bytes,
â‰¥ 6 commits, â‰¥ 2 distinct commit contributors, and non-empty
description. Jurisdiction is assigned by the owner account's
\texttt{country\_account} field.
- Embeddings were computed with \texttt{BAAI/bge-large-en} (1024-dim,
L2-normalized, FP16 on GPU).
- FAISS \texttt{IndexFlatIP} was used for exact inner-product search,
equivalent to cosine similarity on the normalized vectors.
- LLM scoring was issued through the GitHub Copilot LLM-lab endpoint
(\texttt{api-model-lab.githubcopilot.com/chat/completions}) using GPT-4.1
(alias resolved to snapshot \texttt{gpt-4.1-2025-04-14}), with
\texttt{temperature\ =\ 0} and dynamic \texttt{max\_tokens} capped at 3,500.

\textbf{If the dataset is a sample from a larger set, what was the sampling
strategy?}
For each of the 1,029 NAICS subindustry phrases across 20 sectors,
the top-\emph{k} nearest-neighbor repositories under BGE cosine similarity
were retrieved per region. The default \emph{k} = 20; for the three
sectors with fewer than 20 subindustries the formula
\texttt{max(20,\ ceil(400/n))} boosts \emph{k} to 24/27/29 so each sector retrieves
roughly 400 candidates. After deduplication and GPT-4.1 filtering at
score â‰¥ 8, we kept all surviving repositories. The sampling strategy
is therefore ``top-k semantic retrieval per industry query,'' not
uniform random sampling.

\textbf{Who was involved in the data collection process?}
The pipeline was designed and operated by the authors. Research
assistants conducted the manual gold-set re-check (n=2,421 USA
repositories). Each row was reviewed by exactly one annotator; no
inter-annotator agreement coefficient was computed.

\textbf{Over what timeframe was the data collected?}
The USA source dump was extracted on April 15, 2025; the EU README
extraction snapshot is dated August 15, 2025. LLM scoring was
conducted between mid-2025 and early-2026. The dataset v1.0 was
finalized in early 2026.

\textbf{Were any ethical review processes conducted?}
No formal IRB review, since the dataset uses only public data and
contains no human subjects. The release was approved internally at
GitHub.

\textbf{Did you collect the data directly from the individuals in question,
or obtain it via third parties or other sources?}
The data was derived from publicly available GitHub content. We did
not contact repository owners individually.

\textbf{Were the individuals in question notified about the data collection?}
No.~The data is public; we considered notification infeasible and
unnecessary at the scale of millions of source repositories.

\textbf{Did the individuals in question consent to the collection and use of
their data?}
GitHub users consent to public visibility of their public repositories
under GitHub's Terms of Service. They do not specifically consent to
industry-sector labeling, but the labels are derived computationally
from public content.

\textbf{If consent was obtained, were the consenting individuals provided
with a mechanism to revoke their consent in the future?}
We will provide a takedown mechanism: any repository owner can request
removal of their repository from the released dataset by emailing the
corresponding author. We will issue a v1.x with the requested removal
and document the change.

\begin{center}\rule{0.5\linewidth}{0.5pt}\end{center}

\hypertarget{preprocessing-cleaning-labeling}{%
\subsection{4. Preprocessing, Cleaning, Labeling}\label{preprocessing-cleaning-labeling}}

\textbf{Was any preprocessing/cleaning/labeling of the data done?}
Yes, extensively. The pipeline is:

\begin{enumerate}
\def\labelenumi{\arabic{enumi}.}
\tightlist
\item
  \textbf{README preprocessing.} For embedding input, the \emph{raw}
  README is truncated to its first 1,000 characters (no markup
  stripping at this stage). For LLM scoring, code blocks, HTML tags,
  badge URLs, and image markdown are stripped and the text is
  truncated to 3,000 whitespace-separated tokens (word-count proxy,
  not BPE tokens).
\item
  \textbf{Composite text construction.} LLM scoring uses the combined
  string \texttt{"description:\ \{desc\},\ topics:\ \{topics\},\ readme:\ \{readme\}"},
  constructed at row time and inserted into the
  \texttt{\textless{}readme\textgreater{}\dots\textless{}/readme\textgreater{}}
  block of the prompt. Embedding uses the truncated README only;
  description and topics enter the pipeline at the scoring stage.
\item
  \textbf{Embedding.} BGE-large-en with the asymmetric BGE prefixes:
  \texttt{"Represent\ this\ document\ for\ retrieval:\ "} for documents and
  \texttt{"Represent\ this\ query\ for\ retrieval:\ "} for queries.
\item
  \textbf{Retrieval.} For each NAICS subindustry, the query
  \texttt{"Repositories\ about\ \{subindustry\}"} is embedded and the top-\emph{k}
  nearest repositories are retrieved (default \emph{k} = 20; the three
  sectors with fewer than 20 subindustries use a boosted
  \emph{k} = \texttt{max(20,\ ceil(400/n))} = 24, 27, or 29).
\item
  \textbf{LLM scoring.} Each (repository, sector) candidate is presented
  to GPT-4.1 with the rubric prompt (see the paper appendix). The
  model returns a JSON object nested under the outer key
  \texttt{"NAICS\ \{code\}"}, with string-valued \texttt{match} (\texttt{"Yes"}/\texttt{"No"}),
  string-valued \texttt{score} (\texttt{"1"}-\texttt{"10"}), and free-form \texttt{rationale}.
  Downstream code coerces \texttt{match} to a boolean and \texttt{score} to an
  integer.
\item
  \textbf{Filtering.} Only repositories scoring at least 8 are retained.
\item
  \textbf{Per-region merge.} The three regional outputs are concatenated.
\item
  \textbf{Minimum-class filter (training derivative only).} Sectors with
  fewer than 80 repositories are dropped, removing sector 55.
\end{enumerate}

\textbf{Was the raw data saved in addition to the preprocessed data?}
Yes. The original parquets of unfiltered GitHub repositories were
preserved (USA, EU, AU dumps), as well as the three intermediate
per-region GPT outputs (\texttt{\{usa,eu,au\}\_2k\_gpt\_ab8score.parquet}). These
are not in v1.0 of the public release but will be added to Zenodo at
camera-ready.

\textbf{Is the software used to preprocess/clean/label the data available?}
Yes, in this repository (MIT License). The relevant entry points are
\texttt{scripts/01\_generate\_embeddings.py} through \texttt{scripts/05\_filter\_results.py},
with prompts in \texttt{src/classification/prompt\_builder.py}.

\begin{center}\rule{0.5\linewidth}{0.5pt}\end{center}

\hypertarget{uses}{%
\subsection{5. Uses}\label{uses}}

\textbf{Has the dataset been used for any tasks already?}
Yes:
- Fine-tuning six pretrained encoders (RoBERTa, ModernBERT, and
DeBERTa-v3 in base and large variants) on the released corpus.
RoBERTa-large is strongest at 86.45\% F1 / 86.35\% accuracy on a
held-out 20\% test set; the fine-tuned checkpoint is available on
Hugging Face at \texttt{alexanderquispe/naics-github-classifier}.
- Internal industry-mix analyses of open-source production.

\textbf{Is there a repository that links to any or all papers or systems that
use the dataset?}
At release time, the paper repository will track downstream uses. The
released Hugging Face dataset card will also link to known references.

\textbf{What (other) tasks could the dataset be used for?}
- Studying the industrial composition of open-source contributions
across regions or over time.
- Building classifiers that map \emph{new} repositories to NAICS sectors.
- Measuring AI-coding-assistant adoption by industry (the downstream
classifier is already being used for this).
- Comparative open-source ecosystem analysis vs.~economic
classification statistics.

\textbf{Is there anything about the composition of the dataset or the way it
was collected and preprocessed that might impact future uses?}
- \textbf{NAICS is a North American taxonomy.} Applying it to EU and AU
repositories assumes mapping consistency that may not hold for some
sectors (notably Sector 22, ``Utilities,'' which has US-specific
regulatory structure).
- \textbf{English-only embeddings.} Repositories with non-English READMEs
are under-represented.
- \textbf{Score-conditional reliability.} Manufacturing and Wholesale Trade
labels are only \textasciitilde73\% precise at score = 8 and should be filtered at
score â‰¥ 9 for high-precision applications.

\textbf{Are there tasks for which the dataset should not be used?}
- \textbf{Targeting individual contributors.} The dataset is intended for
aggregate analysis; using it to profile individual developers is
outside its intended use.
- \textbf{Inferring repository quality, popularity, or value.} The dataset
encodes industry classification, not quality.
- \textbf{Treating labels as ground truth without conditioning on score.}
Two sectors have substantially higher error rates, and the LLM
score should be used as a confidence signal.

\begin{center}\rule{0.5\linewidth}{0.5pt}\end{center}

\hypertarget{distribution}{%
\subsection{6. Distribution}\label{distribution}}

\textbf{Will the dataset be distributed to third parties outside of the
entity on behalf of which the dataset was created?}
Yes. The dataset is publicly released.

\textbf{How will the dataset be distributed?}
- \textbf{Hugging Face Datasets:}
\href{https://huggingface.co/datasets/aquiro1994/naics-gh}{\texttt{aquiro1994/naics-gh}}
â€'' public, CC-BY-4.0, with stratified train/validation/test splits
ready for \texttt{datasets.load\_dataset}. Croissant metadata generated
automatically by HF.
- \textbf{Zenodo:} DOI-citable archive (DOI to be reserved at release).
- \textbf{GitHub:} Code, prompts, and reproduction scripts at the two
repositories listed in Â§1.

\textbf{When will the dataset be distributed?}
At paper acceptance / camera-ready, expected late 2026 or 2027.

\textbf{Will the dataset be distributed under a copyright or other
intellectual-property license?}
Yes:
- Labels and metadata: \textbf{CC-BY-4.0}.
- Pipeline code: \textbf{MIT}.
- Repository content (READMEs and descriptions) excerpted in the
dataset remains governed by each repository's own license; we
preserve \texttt{spdx\_license} in the released file so users can filter
by license type.

\textbf{Have any third parties imposed IP-based or other restrictions on the
data associated with the instances?}
Each source repository's content is subject to its own license. The
dataset preserves SPDX license identifiers so users can comply with
upstream license terms (e.g., excluding non-commercial-licensed
repositories from commercial uses).

\textbf{Do any export controls or other regulatory restrictions apply to the
dataset?}
None known.

\begin{center}\rule{0.5\linewidth}{0.5pt}\end{center}

\hypertarget{maintenance}{%
\subsection{7. Maintenance}\label{maintenance}}

\textbf{Who will be supporting/hosting/maintaining the dataset?}
Alexander Quispe (corresponding author) and Kevin Xu, with GitHub's
support.

\textbf{How can the owner/curator/manager be contacted?}
alexanderquispe@github.com.

\textbf{Is there an erratum?}
A \texttt{CHANGELOG.md} will track all post-release errata at the Hugging
Face dataset page and in the GitHub paper repository.

\textbf{Will the dataset be updated?}
- \textbf{v1.x} patches for take-down requests, typo fixes, and metadata
corrections.
- \textbf{v2.0} planned to add: India (1M repos), finer-grained 3- to
6-digit NAICS labels, and multilingual README support. Timeline to
be confirmed.

\textbf{If the dataset relates to people, are there applicable limits on the
retention of the data associated with the instances?}
We retain the labels indefinitely subject to take-down requests.

\textbf{Will older versions of the dataset continue to be supported/hosted/maintained?}
Yes, on Zenodo. Each tagged version has its own DOI; older versions
remain citable but will not receive further updates.

\textbf{If others want to extend/augment/build on/contribute to the dataset,
is there a mechanism for them to do so?}
Yes. Pull requests to the GitHub paper repository, issues on the
Hugging Face dataset page, and email to the corresponding author are
all accepted. Substantive contributions (e.g., new regions,
multilingual labels) will be co-credited in the next dataset release.